\documentclass[12pt]{article}

    \usepackage[T1]{fontenc}
    \usepackage{graphicx,epsfig,amsmath,amsfonts,cite}
    \renewcommand{\abstract}{}
\begin{document}

\title{Low-frequency electromagnetic field in a Wigner crystal}
\date{}
\author{Anton Stupka}
 \maketitle

\begin{center} {\small\it Oles Honchar Dnipropetrovs'k National University, Gagarin
ave., 72, 49010 Dnipropetrovs'k, Ukraine\\antonstupka@mail.ru }
\end{center}

\begin{abstract}

Long-wave low-frequency oscillations are described in a Wigner
crystal by generalization of the reverse continuum model for the
case of electronic lattice. The internal self-consistent long-wave
electromagnetic field is used to describe the collective motions
in the system. The eigenvectors and eigenvalues of the obtained system of equations are derived. The velocities of longitudinal and
transversal sound waves are found.

{Keywords:}

 Wigner crystal; self-consistent
electromagnetic field; jellium model;  long wave electromagnetic field; velocity of sound.

\end{abstract}

\section{Introduction}

 The ordered two- and
three-dimensional structures of charged particle systems
 are investigated in numerous recent studies
\cite{[ra],pd,byds}, in particular,  Wigner crystallization in a
quasi-3D electronic system \cite{[pjd]} and the holes in
semiconductors \cite{bfflf} are also actively studied.
 In the paper \cite{[xlm]} the authors dealt
with a two-dimensional Wigner crystal classically as with elastic
medium. In this case the elastic restoring force in the equation
of motion is $\int {{D_{\alpha \beta }}\left( {{\bf{r}} -
{\bf{r'}}} \right)} {u_{2\beta }}\left( {{\bf{r'}},t}
\right){d^2}{\bf{r'}}$ and the external driving force on the total
charge density is given by $ - {n_e}e{\bf{E}},$ where ${D_{\alpha
\beta }}\left( {\bf{r}} \right)$ is the real-space dynamic matrix
tensor, ${{\bf{u}}_2}\left({{\bf{r}},t} \right)$ is the
two-dimensional displacement of the Wigner crystal, ${n_e}$ is its
density, ${\bf{E}}$ is the external electric field.

In this paper we extend the idea presented in~\cite{[xlm]} to
 a three-dimensional
boundless Wigner crystal~\cite{wig,[3]}. We consider a
low-frequency long-wave electromagnetic field in the Wigner
crystal. A usual definition of  an acoustic wave is the following:
an acoustic wave is a joint collective motion of the valence
electrons and ions of lattice in a self-consistent electromagnetic
field \cite[p.~345]{[6]}. However, we have electron lattice and
assume that ions are free as in the jellium model. We consider a
mean electromagnetic field in an acoustic wave and ignore a
dissipation.
\section{Low-frequency waves in a Wigner crystal }

In the case of long acoustic waves, it is possible to rewrite the
elastic restoring force in the form ${C_{\alpha \beta \chi \delta
}}\frac{{{\partial ^2}{u_\delta }}}{{\partial {x_\beta }\partial
{x_\chi }}}$ \cite[p.~152]{[1]}. Here ${C_{\alpha \beta \chi
\delta }}$ is the tensor of the
elastic modules of an electron subsystem. To simplify consideration, we assume that the Wigner crystal is isotropic one. Then using Lame
parameters
it is possible to write down ${C_{\alpha \beta \chi \delta }} =
\lambda{\delta _{\alpha \beta }}{\delta _{\chi \delta }} + \mu\left(
{{\delta _{\alpha \delta }}{\delta _{\beta \chi }} + {\delta
_{\alpha \chi }}{\delta _{\beta \delta }}} \right)$\cite{fs}.   The similar
consideration was given in \cite{s12} for metals. It is known
 that in the case of an ideal crystal without defects of type
of vacancies or interstitials, that is exactly the case we investigate, velocity of
environment points coincides with the derivative of their
displacement with respect to time ${{\bf{v}}_e} =
\partial {\bf{u}}/\partial t$~\cite{[2]}. The electrons oscillate around their equilibrium positions in the lattice.  Short-acting
forces,  correlating separate oscillations, act between them. In the
study of long waves it is possible to unite equation of motion of
elastic subsystem, where short-acting interelectronic
potential $V({u_{n\alpha }})$ appears \cite{[1]}, that gives the elastic
modules, with the long-wave self-consistent electric field.
In the absence of external magnetic field
the relativistic term of
the Lorentz force can be omitted.
Therefore, an equation of
motion for electronic component can be written in the form
    \begin{equation}\label{6}
{\rho _e}\frac{{d^{2} u_\alpha }}{dt^{2}} = \left( {\lambda + \mu}
\right)\frac{{{\partial ^2}{u_\chi }}}{{\partial {x_\alpha
}\partial {x_\chi }}} + \mu\frac{{{\partial ^2}{u_\alpha
}}}{{\partial {x_\chi }\partial {x_\chi }}} - e{n_e}{E_\alpha }.
\end{equation}
A self-consistent electric field satisfies the Maxwell equations
with a hydrodynamic approximation for the current components ${{\bf{j}}_a} = {e_a}{n_a}{{\bf{v}}_a}$, where ${e_a}$
is the corresponding charge~\cite{[4]}. For simplicity all ions are assumed to be identical and
have valence $Z$. We don't consider the piezoelectric or the
magnetic matters. We ignore thermal motion of the ions, and using
the jellium model  we have the the equation of motion for an ionic component
    \begin{equation}\label{3}
{\rho _i}\frac{{d v_{i\alpha }}}{dt} = Ze{n_i}{E_\alpha }.
\end{equation}
 It
means that the Wigner crystal is an elastic electron environment that
contains ``raisins''-ions. Equations (\ref{6}), (\ref{3}) and the
Maxwell equations for a self-consistent electromagnetic field will
allow us to unite the consideration of solid and collective effects.

 We
consider the adiabatic sound waves of small amplitude starting
from the obtained system of equations. For this purpose we linearize  equations of the system near the equilibrium state,
 where
all variables, namely, field strengths and component velocities, are equal
to zero. Then the first  Maxwell equation \cite{[4]} takes the form
\begin{equation}\label{7} \dot{{\bf{E}}} =
c\, \nabla\times{\bf{B}} - 4\pi (Ze{n_{i0}}{{\bf{v}}_i} -
e{n_{e0}}{{\bf{v}}_e}),
\end{equation} where ${n_{a0}}$ is the
equilibrium value of density of the proper particles. The Faraday
equation is linear and, therefore, remains the same
\begin{equation}\label{8}
 \dot{{\bf{B}}} =  -
c\, \nabla\times{\bf{E}}.
\end{equation}                 In this system it is convenient to
pass to the Fourier-components by the following rule
\begin{equation}\label{9} A\left( {{\bf{x}},t} \right) = \smallint
{d^3}k A\left( {{\bf{k}},t } \right){e^{i{\bf{kx}} }}/{(2\pi )^3}.
\end{equation} Then we obtain the system of five linear
homogeneous equations \begin{eqnarray}\label{10} \nonumber
{\bf{\dot E}} = ic[{\bf{k}},{\bf{B}}] - 4\pi
(Ze{n_{i0}}{{\bf{v}}_i} - e{n_{e0}}{{\bf{v}}_e}),\quad {\bf{\dot
B}} = - ic[{\bf{k}},{\bf{E}}], \quad {\bf{\dot u}} = {{\bf{v}}_e},\\
{\rho _{i0}}{{\bf{\dot v}}_i} = Ze{n_{i0}}{\bf{E}},\quad {\rho
_{e0}}{\dot v_{e\alpha }} =  - \left( {\lambda + \mu}
\right){k_\alpha }\left( {{\bf{ku}}} \right) - \mu{u_\alpha }{k^2}
- e{n_{e0}}{E_\alpha }.
\end{eqnarray}
It is observed that the
system for potential and vortical oscillations is divided into two
subsystems. We project all variables on the wave vector ${\bf{k}}$ and  introduce the notations for projections
using the rule  $ {\bf{Ek}}/k =
{E^\parallel }$. Then the system takes the form
\begin{eqnarray}\label{10a}{\dot
E^\parallel } =  - 4\pi (Ze{n_{i0}}v_i^\parallel  -
e{n_{e0}}v_e^\parallel ),\quad {\dot B^\parallel } = 0,\quad
\nonumber\\{\dot u^\parallel } = v_e^\parallel,\quad {\rho
_{i0}}\dot v_i^\parallel  = Ze{n_{i0}}{E^\parallel },\quad {\rho
_{e0}}\dot v_e^\parallel  =  - \left( {\lambda + 2\mu}
\right){u^\parallel }{k^2} - e{n_{e0}}{E^\parallel }.
\end{eqnarray}

It is obvious that the magnetic field does not influence on the
potential motion. It is convenient to introduce the plasma
frequency of corresponding  particles as ${\Omega _a} = \sqrt {4\pi {{\left(
{{e_a}{n_{a0}}} \right)}^2}/{\rho _{a0}}}. $
 We will choose new variables so that the resulting differential equation has no dimensions using the rules
 \[\left( {{E^\parallel }, {u^\parallel },
{v_e}^\parallel , v_i^\parallel } \right) \leftrightarrow \left(
{\frac{{e{E^\parallel }}}{{m{s^2}k}}, {u^\parallel }k,
\frac{{{v_e}^\parallel }}{s}, \frac{{v_i^\parallel M}}{{smZ}}}
\right),\quad t \leftrightarrow tks,\] where  ${s^2} = \frac{{\left(
{\lambda + 2\mu} \right)}}{{{\rho _{e0}}}}.$  In the dimensionless variables the latter system takes the form
\begin{eqnarray}\label{10b} {\dot E^\parallel } = \Omega
_e^2v_e^\parallel  - \Omega _i^2v_i^\parallel,\quad {\dot
u^\parallel } = v_e^\parallel,\quad \dot v_i^\parallel  =
{E^\parallel },\quad \dot v_e^\parallel  =  - {u^\parallel } -
{E^\parallel }.
\end{eqnarray}
In general, the system (\ref{10b}) gives a biquadratic
characteristic equation, but we have to take into account the
smallness of frequency of sound oscillations. As we study long
acoustic waves (with a condition ${\Omega _e} \gg ks$), the
solution with frequency  $\omega  \approx {\Omega _e}$ (here we have taken
into account, that ${\Omega _i} \ll {\Omega _e}$), that
corresponds to high-frequency plasma waves, is not interesting.
The eigenvectors show that electric field, displacement and both
velocities perform coupled oscillations. The eigenvalues
$\Lambda=i\omega$, corresponding to low-frequency branch of
oscillations, are given by
    \begin{equation}\label{15}
\Lambda  =  \pm \frac{i}{2}\sqrt {\Omega _e^2 + \Omega _i^2 + 1 -
     \sqrt {{{\left( {\Omega _e^2 + \Omega _i^2 + 1} \right)}^2} - 4\Omega _i^2} },
\end{equation}
which in limit $\Omega _e \gg ks$ gives the solution
\begin{equation}\label{16}
\Lambda  \approx  \pm i{\Omega _i}/{\Omega _e}.
\end{equation}
Now we return to the dimensional variables and rewrite the frequency of sound
with a dispersion as \begin{equation}\label{20} {\omega ^2} =
\frac{{\left( {\lambda + 2\mu} \right)}}{{{\rho _{i0}}}}{k^2}.
\end{equation}
The velocity of longitudinal sound  \begin{equation}\label{21}
u_s^\parallel  = \sqrt {\frac{{\left( {\lambda + 2\mu}
\right)}}{{{\rho _{i0}}}}}
\end{equation}
 is determined by mass of ions, in analogy to the
ion sound in a two-temperature plasma \cite{[4]}.

Further, we consider the transversal oscillations. For this
purpose we project  equations (\ref{10}) on $\left( {{\delta _{\alpha \beta }}{k^2} -
{k_\alpha }{k_\beta }} \right)$, introducing
notations by the rule $\left( {{\delta _{\alpha \beta }}{k^2} -
{k_\alpha }{k_\beta }} \right){E_\beta } = E_\alpha ^ \bot.$ It is
convenient to pass from induction of the magnetic field to the
new variable ${{\bf{B}}^ \bot } \to \left[
{\frac{{\bf{k}}}{k},{{\bf{B}}^ \bot }} \right] = {\bf{Z}}$
\cite{[ss04]}. In analogy with the longitudinal subsystem we pass to dimensionless variables
using the rules \[\left( {{{\bf{E}}^ \bot },
{\bf{Z}}, {{\bf{u}}^ \bot }, {\bf{v}}_e^ \bot, {\bf{v}}_i^ \bot }
\right) \leftrightarrow \left( {\frac{{e{{\bf{E}}^ \bot
}}}{{m{s^2}k}}, \frac{{e{\bf{Z}}}}{{m{s^2}k}}, {{\bf{u}}^ \bot }k,
\frac{{{\bf{v}}_e^ \bot }}{s}, \frac{{{\bf{v}}_i^ \bot M}}{{smZ}}}
\right),\quad t \leftrightarrow tks,\quad c \leftrightarrow c/s,\] where  ${s^2} = \frac{\mu}{{{\rho _{e0}}}}.$ In the dimensionless variables we
have the system
\begin{eqnarray}\label{22} \nonumber{{\bf{\dot E}}^ \bot } = ic{\bf{Z}} +
\Omega _e^2{\bf{v}}_e^ \bot  - \Omega _i^2{\bf{v}}_i^ \bot,\quad
{\bf{\dot Z}} = ic{{\bf{E}}^ \bot },\\ {{\bf{\dot u}}^ \bot } =
{\bf{v}}_e^ \bot,\quad {\bf{\dot v}}_i^ \bot = {{\bf{E}}^ \bot
},\quad {\bf{\dot v}}_e^ \bot =  - {{\bf{u}}^ \bot } - {{\bf{E}}^
\bot }.
\end{eqnarray}
The  eigenvectors show that all physical variables perform coupled
oscillations. The eigenvalues that correspond to low-frequency
branch of oscillations are
    \begin{equation}\label{15t}
\Lambda  =  \pm \frac{i}{2}\sqrt {\Omega _e^2 + \Omega _i^2 + 1 +
{c^2} -
    \sqrt {{{\left( {\Omega _e^2 + \Omega _i^2 + 1 + {c^2}} \right)}^2} - 4\left( {\Omega _i^2 + {c^2}} \right)} }.
    \end{equation}
In the limit $\Omega _e \gg kc$  for the condition ${\Omega _i} \gg
kc$ this gives the transversal sound
\begin{equation}\label{16t}
\Lambda  \approx  \pm i{\Omega _i}/{\Omega _e}.
\end{equation}
 However, in the area  ${\Omega _i} \ll
kc\ll{\Omega _e}$ there is the quadratic dispersion $\Lambda \approx
\pm ic/{\Omega _e}.$

Returning to the dimensional variables we obtain low-frequency mode
\begin{equation}\label{lt} {\omega ^2} = \frac{\mu}{{{\rho
_{i0}}}}{k^2} + {c^2}\frac{\mu}{{{\rho _{e0}}\Omega _e^2}}{k^4}.
\end{equation}
Equation (\ref{lt}) demonstraits non-linearity for middle wavelengths (see
Apendix). Expressions  (\ref{15t}) and (\ref{lt}) approach to the sound, in supposition
${\Omega _i} \gg kc$, with dispersion
\begin{equation}\label{27} {\omega ^2} = \frac{\mu}{{{\rho
_{i0}}}}{k^2}.
\end{equation}
 Therefore, the velocity of transversal sound is
\begin{equation}\label{28}
u_s^ \bot  = \sqrt {\frac{\mu}{{{\rho _{i0}}}}}.
\end{equation}
It is easy to see that  the velocities satisfy the well-known requirement
$u_s^\parallel /u_s^ \bot  = \sqrt {2 +
\lambda/\mu}
> \sqrt 2$~\cite[p.~125]{[2]}.

Now, starting from systems (\ref{10b}) and
(\ref{22}), it is easy to show that expressing velocities of
subsystems of charges  from equations of motion through the fields
and substituting them in the corresponding wave equation  obtained from the Maxwell
equations, the sound oscillations reduce to oscillations
of the electric field in the crystal. The quanta of the obtained
sound oscillations are the quanta of the electromagnetic field in
the crystal. That gives correct statistics for phonons, as bosons
with zero chemical potential. Therefore, it is possible in standard
way to introduce operators of annihilation and creation of phonons as
quanta of the electromagnetic field after the decomposition of the
vector potential of electromagnetic field on plane waves ${\hat
A_n}\left( x \right) = c{\sum\limits_{k\alpha } {\left(
{\frac{{2\pi \hbar }}{{\omega V}}} \right)} ^{{1 \mathord{\left/
 {\vphantom {1 2}} \right.
 \kern-\nulldelimiterspace} 2}}}\left( {{c_{k\alpha }} + c_{ - k,\alpha }^ + } \right){e_{k\alpha n}}{e^{ikx}}$
 \cite{[7]}, where the index $\alpha $ denotes longitudinal and two transversal polarizations.

\section{Conclusion}

Using the jellium (continuum) model of a solid by introducing an
elasticity of electronic subsystem for a Wigner crystal, we have found
the longitudinal and two transversal low-frequency oscillation
brunches. We have shown that these sound oscillations in a Wigner
crystal can be considered as  the coupled waves of the electromagnetic field and
charges which it is possible to study as waves of the
electromagnetic field in an environment. It gives phonons with
necessary statistics after the standard quantization of the field.

\section*{Appendix}

For numerical estimation we take the average inter-particle
spacing of order $a\sim 100 a_{B}$ \cite{drttn}, where
$a_{B}\approx 5.29\cdot10^{-9} \, cm$ is the Bohr radius. Then an
electronic plasma frequency is $\Omega _e \sim 1.5\cdot10^{14}\,
s^{-1} $ and let an ionic one be $\Omega _i \sim\Omega _e/100 $.
We estimate shear modulus using Coulomb interaction as $\mu\sim
e^{2}/a^{4}\sim 3 \cdot10^{6}\,g/(cm\cdot s^{2})$
 and from (\ref{28}) obtain $u_s^
\bot \sim 2\cdot 10^{5}\,cm/s$. In this supposition we get the
numerical estimation for low-frequency mode (\ref{lt})
\[ {\omega } =2 \sqrt{10^{10}{k^2} +4\cdot 10^{6}{k^4}}\]
and plot this dispersion dependence.
\begin{figure}[h]
\centering
\includegraphics[width=8cm]{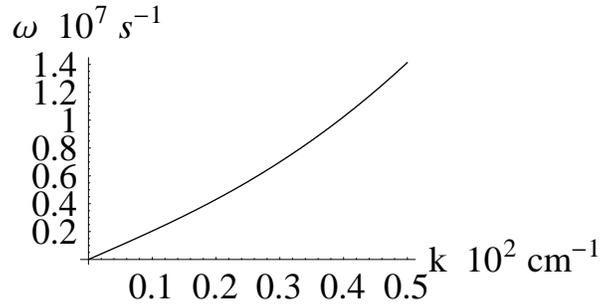}
\caption{Low-frequency transversal mode.} \label{graf}
\end{figure}

Figure \ref{graf} shows the transition from linear sound
dispersion for small wavevectors to nonlinear dispersion for
middle ones.

\subsection*{Aknowledgement} The author would like to thank unknown referee, Drs. O.~Vaneeva and O.~Kovalchuk for useful comments and constructive suggestions.

\end{document}